\documentclass[%
reprint,
amsmath,amssymb,
aps,
]{revtex4-1}
\usepackage{graphicx}
\usepackage{epstopdf}
\usepackage{color}
\begin{document}
	
\preprint{APS}

\title{Optical Signature of Flat Bands in Topological Hourglass Semimetal Nb$_{3}$SiTe$_{6}$}% Force line breaks with \\
\author{Shize Cao$^{1,2,\sharp}$}
\author{Cuiwei Zhang$^{2,\sharp}$}
\author{Yueshan Xu$^{2,\sharp}$}
\author{Jianzhou Zhao$^{3,\S}$}
%\email{jzzhao@swust.edu.cn}
\author{Youguo Shi$^{2,4}$}
\author{Yun-Ze Long$^{1,\dagger}$}
%\email{yunze.long@qdu.edu.cn}
\author{Jianlin Luo$^{2,4}$}
\author{Zhi-Guo Chen$^{2,4,*}$}
%\email{zgchen@iphy.ac.cn}
\affiliation{$^{1}$Collaborative Innovation Center for Nanomaterials $\&$ Devices, College of Physics, Qingdao University, Qingdao 266071, China\\
	$^{2}$Beijing National Laboratory for Condensed Matter Physics, Institute of Physics, Chinese Academy of Sciences, Beijing 100190, China\\
	$^{3}$Co-Innovation Center for New Energetic Materials, Southwest University of
	Science and Technology, Mianyang 621010 Sichuan, China\\
	$^{4}$Songshan Lake Materials Laboratory, Dongguan, Guangdong 523808, China
}

	 %Corresponding authors.Email:zgchen@iphy.ac.cn; yunze.long@qdu.edu.cn;jzzhao@swust.edu.cn
	
	%\altaffiliation[] 
	{}%Lines break automatically or can be forced with \\

	\begin{abstract}
	
		Flat electronic bands in condensed matter provide a rich avenue for exploring novel quantum phenomena. Here, we report an optical spectroscopy study of a topological hourglass semimetal Nb$_{3}$SiTe$_{6}$ with the electric field of the incident light parallel to its crystalline {\textit{ab}}-plane. The {\textit{ab}}-plane optical conductivity spectra of Nb$_{3}$SiTe$_{6}$ single crystals exhibit a remarkable peak-like feature around 1.20 eV, which is mainly contributed by the direct optical transitions between the two \textit{ab-initio}-calculation-derived flat bands along the momentum direction Z-U. Our results pave the way for investigating exotic quantum phenomena based on the flat bands in topological hourglass semimetals.
	
	%	\begin{description}
		%	\item[Keywords]
		%	 Flat band; Topological Semimetal; Optical spectroscopy
	%	\end{description}
	\end{abstract}
	

	\maketitle
	
%	\section*{Introduction}\vspace{-1mm}
	Flat electronic bands, which represent dispersionless bands with divergent density of state (DOS), have been attracting great interest in condensed matter community due to their crucial roles in the emergence of various quantum phenomena, such as fractional quantum Hall effect \cite{1,2,3,4,5}, superconductivity \cite{6,7,8,10,11,12,M,14}, and ferromagnetism \cite{18,19,20,21,22,23,24}. Several electron systems, including kagome-lattice compounds \cite{C-1,C-2,25,26,27,28,29,30,31,32,33,34,35,36,37,38,39,40,41,42,43}, twisted bilayer graphene \cite{a}, Weyl semimetals \cite{44,45,46,47}, and heavy-fermion materials \cite{49}, possess flat electronic bands \cite{C-3,50,51}. An important question is whether flat bands can be observed in a broader class of electron systems.
	
	A van der Waals material Nb$_{3}$SiTe$_{6}$ shows a topological hourglass semimetal phase with the hourglass-type band structure protected by the nonsymmorphic space group symmetry \cite{52,53,54}. Furthermore, the signatures of weak topological insulator were identified in Nb$_{3}$SiTe$_{6}$  \cite{55}. Electrical transport study indicates that electron coherence is unexpected enhanced in few-layer Nb$_{3}$SiTe$_{6}$ owing to the quantum-confinement-induced suppression of electron-phonon interactions \cite{56}. Moreover, Nb$_{3}$SiTe$_{6}$ thin flakes exhibit Shubnikov-de Haas oscillations with nontrivial $\pi$-Berry phase \cite{57}, which implies the nontrivial topological nature of its electronic bands. It was theoretically predicted that the Berry curvature dipole in Nb$_{3}$SiTe$_{6}$ should produce nonlinear Hall effect \cite{58}. Additionally, the device made of Nb$_{3}$SiTe$_{6}$ has a quite high thermoelectric power and a very large power factor \cite{59}. Recently, linearly polarized infrared spectroscopy measurements of Nb$_{3}$SiTe$_{6}$ show the excitations around 0.15 eV and 0.28 eV \cite{60}, which were  ascribed to the fingerprints of van Hove singularities in its electronic band structure. However, the optical signature of the flat bands in Nb$_{3}$SiTe$_{6}$ has rarely been observed by experiments.
	
	\begin{figure*}
		\centering
		\includegraphics[width=16.5 cm]{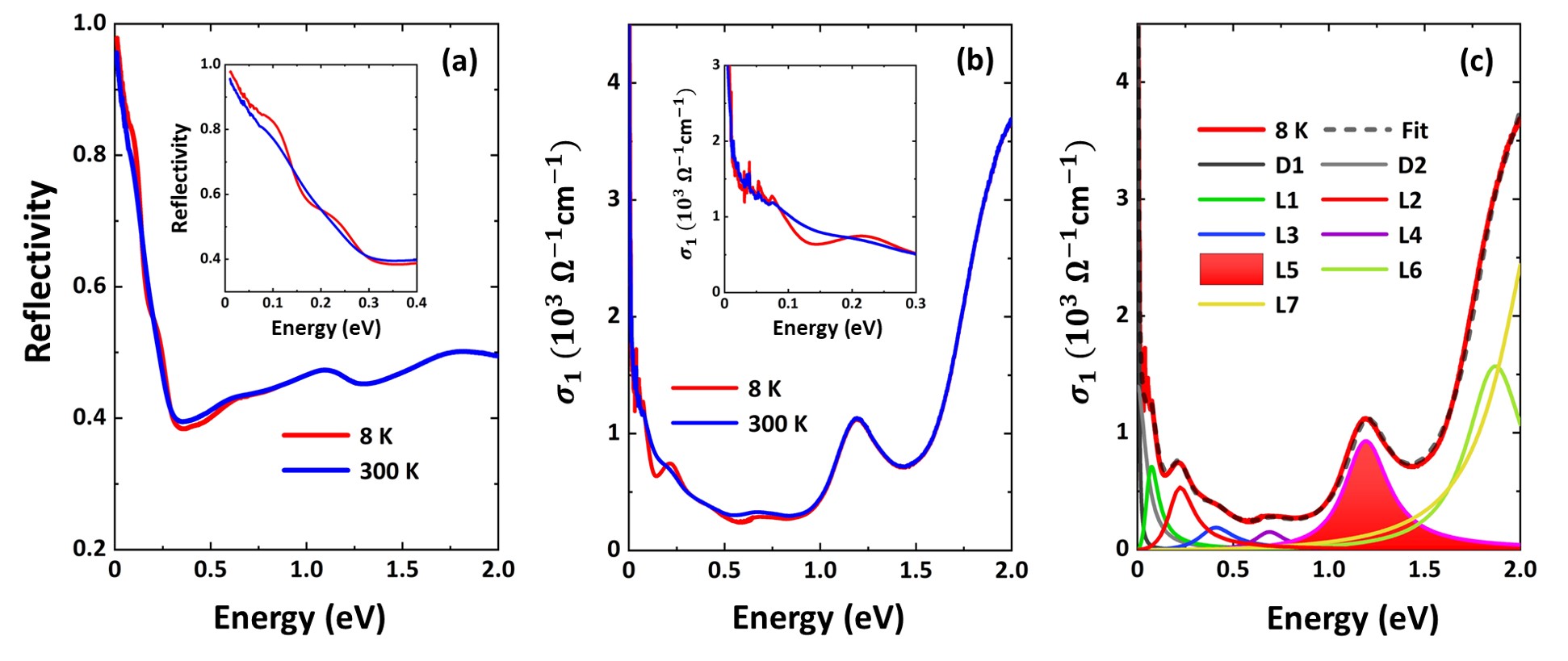}
		\caption{(a) Reflectance spectra of Nb$_{3}$SiTe$_{6}$ measured with the electric field of the incident light parallel to the crystalline {\textit{ab}}-plane at two typical temperatures \textit{T} = 8K and 300K in the energy up to 2 eV. The inset in (a) shows the reflectance spectra in the energy range up to 0.4 eV. (b) Real part $\sigma_{1} (\omega)$ of the {\textit{ab}}-plane optical conductivity of Nb$_{3}$SiTe$_{6}$ at two typical temperatures \textit{T} = 8K and 300K in the energy up to 2 eV. The inset in (b) displays the $\sigma_{1} (\omega)$ in the energy range up to 0.4 eV. (c) Drude-Lorentz fit to the {\textit{ab}}-plane $\sigma_{1} (\omega)$ of Nb$_{3}$SiTe$_{6}$ at \textit{T} = 8K. D$_1$ and D$_2$ represent the two Drude components. L$_1$--L$_7$ are the Lorentzian components. The Lorentzian component L$_5$ shaded in red color indicates the distinct peak-like feature around 1.20 eV.}
	\end{figure*}
	
	Optical spectroscopy is an efficient experimental technique for probing interband transitions from occupied to empty states \cite{61,62,64,65,66} and thus enables us to search for the optical signature of the flat electronic bands in Nb$_{3}$SiTe$_{6}$. Here, we present the optical reflectance measurements of the Nb$_{3}$SiTe$_{6}$ single crystals over broad photon energies with the directions of the electric fields of the incident light parallel to the newly cleaved surfaces (i.e., crystalline \textit{ab}-plane with the in-plane lattice parameters \textit{a} = 11.5 \AA, \textit{b} = 6.3 \AA). Our optical reflectance measurements reveal a remarkable peak-like feature around 1.20 eV in the real part (i.e., $\sigma_{1} (\omega)$) of the optical conductivity spectra which are induced by the electric fields of the incident light parallel to the \textit{ab}-plane of Nb$_{3}$SiTe$_{6}$. Moreover, two flat bands obtained by \textit{ab-initio} calculations are separately located at -- 0.21 eV and 0.99 eV along the momentum direction Z-U and has the energy distance $\sim$ 1.20 eV, which is consistent with the energy postion of the remarkable peak-like feature around 1.20 eV. Thus, the peak-like feature around 1.20 eV in the $\sigma_{1} (\omega)$, which arises primarily from the direct optical transitions between the two flat bands along Z-U, can be assigned as an optical signature of the flat bands in Nb$_{3}$SiTe$_{6}$.
	
	\vspace{0mm}
	
%	\section*{Results and Discussions}
	The Nb$_{3}$SiTe$_{6}$ single crystals were grown by a chemical vapor transport method \cite{56}. High-purity elemental Nb, Si, and Te with a molar ratio of 3:2:6 were putted into a quartz tube under high vacuum. The quartz tube was placed in the two-zone furnace where the temperatures of the hot and cool sides were set as 1223 K and 1123 K, respectively.

	The optical reflectance measurements in the temperature range from 300 K to 8 K were performed on a Bruker Vertex 80V Fourier transform spectrometer. The reflectance spectra (i.e., $R(\omega)$) of the freshly cleaved {\textit{ab}}-plane of the Nb$_{3}$SiTe$_{6}$ single crystals were obtained using an \textit{in-situ} gold and aluminum overcoating technique. The Kramers-Kronig transformation of the $R(\omega)$ was used to get the $\sigma_{1} (\omega)$ of Nb$_{3}$SiTe$_{6}$.

	Figure 1(a) displays the {\textit{ab}}-plane $R(\omega)$ of the Nb$_{3}$SiTe$_{6}$ single crystals at two typical temperatures \textit{T} = 8 K and 300 K in the energy range up to 2 eV. As shown in the inset of Fig. 1(a), the $R(\omega)$ at \textit{T} = 8 K at energies lower than 30 meV is higher than the $R(\omega)$ at \textit{T} = 300 K and approaches to unity, which exhibits a good metallic response. Beside the metallic optical response at low energies, several hump-like features can be clearly observed at high energies in the $R(\omega)$ at \textit{T} = 8 K.

	To study the nature of the hump-like features present in the $R(\omega)$ of Nb$_{3}$SiTe$_{6}$, we obtained the $\sigma_{1} (\omega)$ by the Kramers-Kronig transformation of the $R(\omega)$. Figure 1(b) depicts the $\sigma_{1} (\omega)$ at \textit{T} = 8 K and 300 K. In the inset of Fig. 1(b), upturn-like features---Drude components coming from the optical response of the free carriers can be observed at energies lower than 50 meV in the $\sigma_{1} (\omega)$ at \textit{T} = 8 K and 300 K, respectively, which is in agreement with the metallic behavior of the $R(\omega)$. Moreover, corresponding to the hump-like features in the $R(\omega)$, a series of peak-like features are present in the $\sigma_{1} (\omega)$.
	
	To identify the energy positions of the peak-like features in Fig. 1, we fit the $\sigma_{1} (\omega)$ based on a standard Drude-Lorentz model:
	\begin{equation}
		\sigma_{1}=\frac{2\pi}{Z_{0}}\frac{\omega_{D}^2\Gamma_{D}}{\omega^2+\Gamma_{D}^2}+\sum_{j}^{ }\frac{2\pi}{Z_{0}}\frac{S_{j}^2\omega^2\Gamma_{j}}{(\omega_{j}^2-\omega^2)^2+\omega^2\Gamma_{j}^2}
	\end{equation}
	where $Z_{0}$ = 377 $\Omega$ is the electric impedance, $\omega_{D}$ and $\Gamma_{D}$ is the plasma energy and the relaxation rate of free carriers, respectively, while $\omega_{j}$, $S_{j}$, and $\Gamma_{j}$ are the resonance energy, the mode strength and the damping of each Lorentzian term, respectively. The first term in Eq. (1) represents the Drude component. The rest of the terms in Eq. (1) are the Lorentzian components which can be used to describe the optical responses of interband transitions. Figure 1(c) displays the Drude-Lorentz fit to the $\sigma_{1} (\omega)$ at \textit{T} = 8 K (see the fitting parameters in Table 1). Therein, the energy positions (i.e., 0.22 eV, 0.41 eV, and 2.13 eV) of the Lorentzian peaks L$_2$, L$_3$ and L$_7$ are close to those (i.e., 0.28 eV, 0.41 eV, and 2.00 eV) of the Lorentian peaks previously measured by linearly polarized infrared spectroscopy. It is worth noticing that a remarkable peak-like feature, which was fitted by the Lorentzian peak L$_5$ shaded in red color, appears around 1.20 eV in Fig. 1(c). However, the peak-like feature around 1.20 eV in the $\sigma_{1} (\omega)$ obtained by our optical reflectance measurements cannot be well resolved in the previously reported $\sigma_{1} (\omega)$ gotten by linearly polarized infrared spectroscopy. The electric fields (\textbf{\textit{E}} // \textit{ab}-plane) of the incident light utilized in our optical reflectance study not only can excite the inter-band transitions coupled with the electric fields (\textbf{\textit{E}} // \textit{a}-axis and \textbf{\textit{E}} // \textit{b}-axis) of the incident light used in the previously linearly polarized infrared study but also can induce the inter-band transitions coupled with the electric fields of the incident light with the direction between the \textit{a}-axis and the \textit{b}-axis. Thus, the peak-like feature around 1.20 eV in the $\sigma_{1} (\omega)$ obtained by our optical reflectance measurements, which is invisible in the previously reported $\sigma_{1} (\omega)$, may arise from the inter-band transitions coupled with the electric fields of the incident light with the directions between the \textit{a}-axis and the \textit{b}-axis.
	
	\begin{figure*}[t]
		\centering
		\includegraphics[width=13 cm]{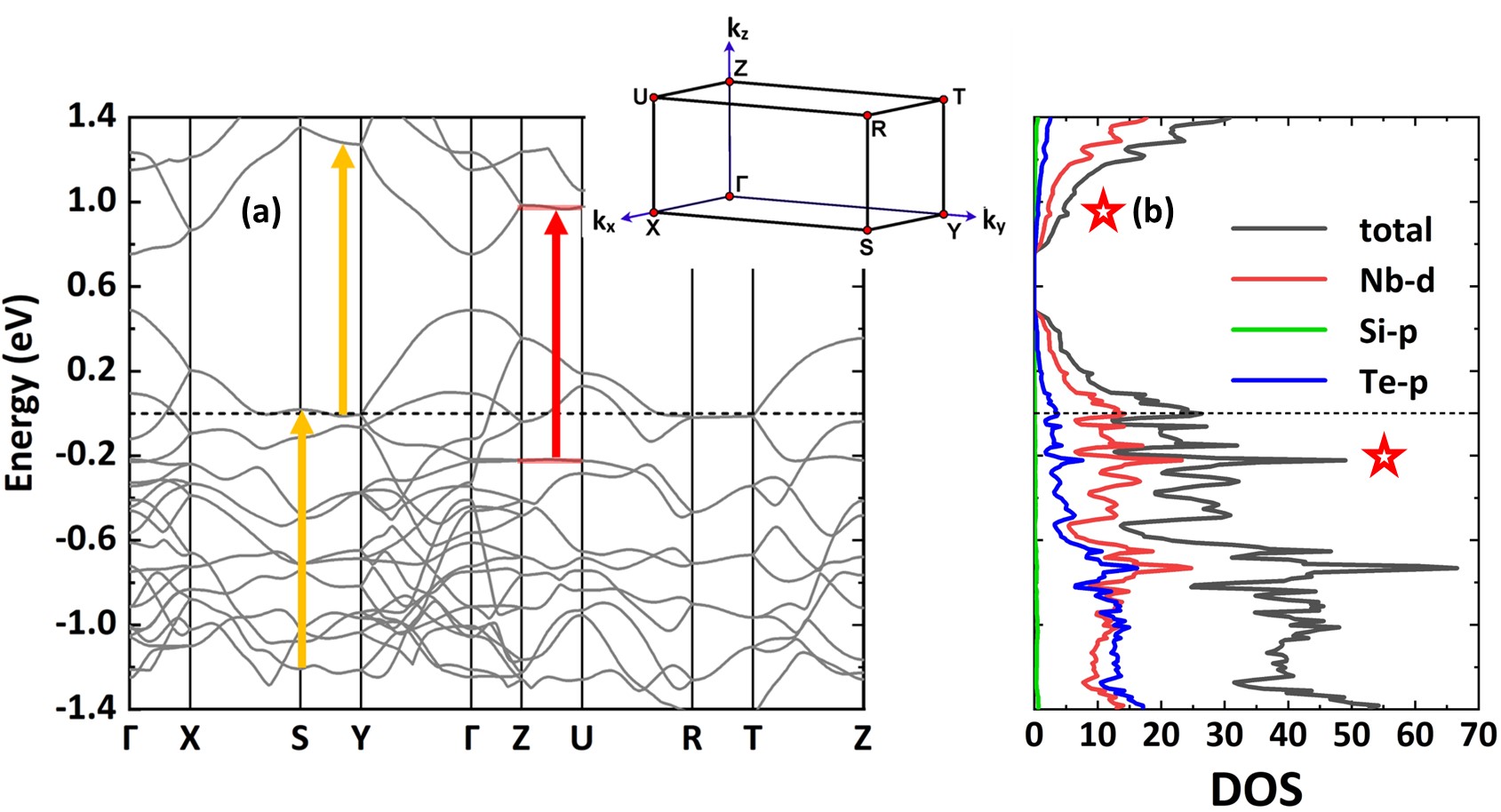}\vskip 2mm
		\caption{(a) Electronic band structure of Nb$_{3}$SiTe$_{6}$ obtained by \textit{ab-initio} calculations.The red vertical arrow indicates the the direct optical transitions between the two flat bands. The inset of (a) shows the high symmetry directions of the first Brillouin zone of Nb$_{3}$SiTe$_{6}$. The first Brillouin zone here corresponds to the in-plane lattice parameters \textit{a} = 11.5 \AA, \textit{b} = 6.3 \AA, and the lattice parameter along the interlayer direction \textit{c} = 13.9 \AA. (b) Total density of states and partial density of states of the electronic bands with Nb 4d, Si 2p and Te 5p orbital characters.}
	\end{figure*}

	We checked whether the peak-like feature around 1.20 eV is intimately associated with the optical absorptions of excitons and polarons.The Drude-Lorentz fit to the $\sigma_{1} (\omega)$ at \textit{T} = 8 K indicates that the plasma energy of the total free charge carriers in Nb$_{3}$SiTe$_{6}$ $\omega_{D}^T$ = $\sqrt{\omega_{D1}^2+\omega_{D1}^2}$ = 1000 meV, which is higher than those in some good metals such as Co$_{3}$Sn$_{2}$S$_{2}$ with the $\omega_{D}$ = 258 meV \cite{62}, and BaFe$_{2}$As$_{2}$ with the $\omega_{D}$ = 582 meV \cite{64,65,66}. It is well known that in good metals, the high concentration of free charge carriers means a significant screening of the Coulomb interactions between the electron and hole of an exciton, which leads to the absence of the excitonic feature. Thus, the peak-like feature around 1.20 eV in the $\sigma_{1} (\omega)$ of Nb$_{3}$SiTe$_{6}$ with the high charge carrier density and the significant Coulomb screening is unlikely to arise from an exciton absorption \cite{67,68,69,70,71,72}. Furthermore, it was reported that if the plasma energy is not higher than the longitudinal-optical-phonon energy, the longitudinal-optical phonon would be insufficiently screened by charge free carriers, which results in the observation of the polaron feature \cite{73,74,75}. However, the large plasma energy (i.e., 1000 meV) in Nb$_{3}$SiTe$_{6}$ is generally higher than the longitudinal-optical phonon in this topological hourglass semimetal, which indicates that the longitudinal-optical phonon in Nb$_{3}$SiTe$_{6}$ should be significantly screened by the charge free carriers and suggests that the peak-like feature around 1.20 eV should not come from a polaron absorption.

	To identify the origin of the peak-like feature around 1.20 eV in the $\sigma_{1}(\omega)$ here, we performed \textit{ab-initio} calculations on the electronic band structure and DOS of Nb$_{3}$SiTe$_{6}$. The electronic band structure of Nb$_{3}$SiTe$_{6}$ was calculated via the density functional theory, within the generalized gradient approximation (GGA) of Perdew-Burke-Ernzerhof (PBE) realization, as implemented in the Vienna ab initio Simulation Package (VASP). The projector augmented wave (PAW) pseudopotentials were adopted for the calculations. Electrons belonging to the outer atomic configuration were treated as valence electrons, here corresponding to 11 electrons in Nb (4p$^6$4d$^4$5s$^1$), 6 electrons in Te (5s$^2$5p$^4$) and 4 electrons in Si (3s$^2$3p$^2$). The kinetic energy cutoff was fixed to 400 eV, and a $\Gamma$-center grid of 5$\times$10$\times$4 \textit{k}-points for self-consistency. The energy convergence accuracy set to 10$^{-7}$ eV.

	\begin{table}
		\begin{centering}
			\caption{Parameters of the Drude-Lorentz fit to the $\sigma_{1} (\omega)$ at \textit{T} = 8 K.}
			\begin{tabular}{cccccc}
				\hline\hline
				j   &   $\omega_{j}$(meV)   &   $\Gamma_{j}$(meV)   &   $S_{j}$(meV)    &    $\omega_{D}$(meV)   &    $\Gamma_{D}$(meV) \\
				\hline
				D$_{1}$ & - & - & - & 600 & 46\\
				D$_{2}$ & - & - & - & 800 & 53\\
				L$_{1}$ & 74 & 82 & 710 & - & -\\
				L$_{2}$ & 224 & 160 & 860 & - & -\\
				L$_{3}$ & 410 & 82 & 600 & - & -\\
				L$_{4}$ & 700 & 190 & 500 & - & -\\ 
				L$_{5}$ & 1200 & 280 & 1465 & - & -\\
				L$_{6}$ & 1860 & 312 & 2240 & - & -\\
				L$_{7}$ & 2130 & 580 & 3770 & - & -\\
				\hline\hline
			\end{tabular}

		\par\end{centering}
	\end{table}

	Figure 2(a) displays the calculated electronic bands along high symmetry directions in the energy range from -- 1.40 eV to 1.40 eV. It is worth noticing that along the momentum direction Z-U, two flat bands indicated by the red lines in Fig. 2(a) are located at -- 0.21 eV and 0.99 eV, respectively. The energy distance of these two flat bands is $\sim$ 1.20 eV, which is consistent with the energy postion of the remarkable peak-like feature around 1.20 eV in the $\sigma_{1} (\omega)$. Moreover, Fig. 2(b) shows that the calculated DOS has two peaks indicated by the red stars at the energy positions (i.e., -- 0.21 eV and 0.99 eV) of the two flat bands. Although the DOS peak at 0.99 eV is relatively weak, the DOS peak at -- 0.21 eV is quite strong. The quite strong DOS peak at -- 0.21 eV and the relatively weak DOS peak at 0.99 eV are expected to result in a large value of the joint DOS. As is well known, the absorption intensity of the optical transitions in solids is determined by the joint DOS of the occupied state below Fermi energy and the empty state above Fermi energy \cite{61}, so the two DOS peaks at -- 0.21 eV and 0.99 eV should lead to a peak-like structure at $\sim$ 1.20 eV in the optical absorption spectra of Nb$_{3}$SiTe$_{6}$. Thus, the distinct peak-like feature around 1.20 eV in the $\sigma_{1} (\omega)$ can be assigned as the direct optical transitions between the two flat bands in Nb$_{3}$SiTe$_{6}$ (see the red vertical arrow in Fig. 2(a)), which host the DOS peaks and are separately located at -- 0.21 eV and 0.99 eV. Furthermore, the two direct optical transitions between the non-flat bands along the momentum direction S-Y, which are indicated by the yellow verticle arrows in Fig. 2(a), have the energies of $\sim$ 1.20 eV and should contribute to the formation of the peak-like feature around 1.20 eV in the $\sigma_{1} (\omega)$ as well. However, the DOSs of the non-flat bands related to the direct optical transitions along the direction S-Y should be lower than those of the two flat bands at -- 0.21 eV and 0.99 eV along the direction Z-U. Therefore, the remarkable peak-like feature around 1.20 eV is mainly contributed by the direct optical transitions between the two flat bands along the momentum direction Z-U. If these two flat bands are separately tuned to be at the Fermi energy via electron-like and hole-like charge-carrier doping, the DOS at the Fermi energy would be quite high, which may induce superconductivity in this topological hourglass semimetal \cite{6,8,M}.{It was previously reported that in Mo$_{5}$Si$_{3}$, its flat band along X-M originally far away from the Fermi level moves towards the Fermi level after partially replacing the silicon atoms by phosphorus atoms \cite{M}. In Mo$_{5}$Si$_{{3-x}}$P$_{{x}}$ (x $\simeq$ 1.3), its flat band with a negligible modification of the energy-momentum dispersions is located at the Fermi level. Meanwhile, their superconducting quantities have been significantly changed.} In addition, we have compared the other peak-like features L$_2$, L$_3$, L$_4$, L$_6$ and L$_7$ with the calculated band structure and DOS. The peak-like features L$_2$, L$_3$, L$_4$, L$_6$ and L$_7$ can be assigned as the optical transitions between the bands with high DOS (see the red vertical arrows in Supplemental Fig. 1(a)-(e)).
	
%	\section*{Conclusion}
	In summary, we have studied the the nature of the distinct peak-like feature around 1.20 eV in the $\sigma_{1} (\omega)$ of Nb$_{3}$SiTe$_{6}$ using optical spectroscopy with the electric field of the incident light parallel to the crystalline {\textit{ab}}-plane. The peak-like feature around 1.20 eV arises primarily from the direct optical transitions between the two flat bands with the high DOSs and the energy distance $\sim$ 1.20 eV along the momentum direction Z-U and can serve as an optical signature of the flat bands in topological hourglass semimetal Nb$_{3}$SiTe$_{6}$.
	
	\vspace{2mm}
%	\section*{Acknowledgment} \vspace{-5pt}
	\textit{Acknowledgment} This work was supported by the Guangdong Basic and
	Applied Basic Research Foundation (Projects No. 2021B1515130007), the National Natural Science 	Foundation of China (Grant No. U21A20432 and 52273077), the National Key
	Research and Development Program of China (Grant No. 2022YFA1403800), the strategic Priority Research Program of Chinese Academy of Sciences (Project No. XDB33000000), and the Synergetic Extreme Condition User Facility (SECUF,~{https://cstr.cn/31123.02.SECUF})--Infrared Unit in THz and Infrared Experimental Station.
	\vspace{3mm}
	
	$*$ zgchen@iphy.ac.cn;
	
	$\dagger$ yunze.long@qdu.edu.cn; 
	
	$\S$ jzzhao@swust.edu.cn
	
	$\sharp$ These author contributed equally to this work.

	%\vskip 4mm
	
	%\fl{4}\centerline{\includegraphics[width=0.98\linewidth]{Figure2}}
	
	%\vskip 2mm
	
	%\figcaption{8}{2}{(a) Electronic band structure of Nb$_{3}$SiTe$_{6}$ obtained by \textit{ab-initio} calculations.The red vertical arrow indicates the the direct optical transitions between the two flat bands. The inset of (a) shows the high symmetry directions of the first Brillouin zone of Nb$_{3}$SiTe$_{6}$. (b) Total density of states and partial density of states of the electronic bands with Nb 4d, Si 2p and Te 5p orbital characters.}
	%\vskip 4mm
	%\medskip

\end{document}